\documentclass[conference]{IEEEtran}
\usepackage{cite}
\usepackage{amsmath,amssymb,amsfonts}
\usepackage{graphicx}
\usepackage{textcomp}
\usepackage{xcolor}
\usepackage{fancyhdr}
\usepackage[hyphens]{url}
\usepackage{multirow}
\usepackage{authblk}
\usepackage{graphicx}
\graphicspath{{fig/}}
\usepackage{xspace}
\usepackage{subfig}
\usepackage{xcolor}
\usepackage{amsmath}
\usepackage{array,multirow}
\usepackage[ruled,vlined,noend]{algorithm2e}
\usepackage{stfloats}
\usepackage[bottom,flushmargin]{footmisc}
\usepackage[colorinlistoftodos,prependcaption]{todonotes}
\usepackage{soul}
\usepackage[normalem]{ulem}

\newcommand{\mcut}[2][]{}

\newcommand{\scratch}[1]{}

\usepackage{makecell}                                                                  
\usepackage{xcolor, colortbl}   
\usepackage{tabu}  
\usepackage{array,booktabs,multirow}

\newcolumntype{C}{>{\centering\arraybackslash}p{8em}}
\DeclareMathSymbol{\minus}{\mathord}{operators}{"2D}

\definecolor{mygreen}{RGB}{0,90,0}

\newcommand{\fig}[1]{Figure~\ref{#1}\xspace}

\newcommand{\meadd}[1]{\textcolor{black}{#1}}
\newcommand{\medel}[1]{\textcolor{purple}{}}

\newcommand{\badd}[1]{\textcolor{black}{#1}}

\newcommand{\hpcacut}[1]{}

\def\BibTeX{{\rm B\kern-.05em{\sc i\kern-.025em b}\kern-.08em
    T\kern-.1667em\lower.7ex\hbox{E}\kern-.125emX}}

\fancypagestyle{firstpage}{
  \fancyhf{}

}

\title{Accelerating Bandwidth-Bound Deep Learning Inference with Main-Memory Accelerators} 
\author[1]{Benjamin Y. Cho, Jeageun Jung, and Mattan Erez}
\affil[1]{The University of Texas at Austin}
\affil[ ]{\textit {\{bjcho,jeageunjung,mattan.erez\}@utexas.edu}}

\begin{document}
\maketitle
\thispagestyle{firstpage}
\pagestyle{plain}


\begin{abstract}
~\\
DL inference queries play an important role in diverse internet services and a large fraction of datacenter cycles are spent on processing DL inference queries. Specifically, the matrix-matrix multiplication (GEMM) operations of fully-connected MLP layers dominate many inference tasks. We find that the GEMM operations for datacenter DL inference tasks are memory bandwidth bound, contrary to common assumptions: (1) strict query latency constraints force small-batch operation, which limits reuse and increases bandwidth demands; and (2) large and colocated models require reading the large weight matrices from main memory, again requiring high bandwidth without offering reuse opportunities. We demonstrate the large potential of accelerating these small-batch GEMMs with processing in the main CPU memory. We develop a novel GEMM execution flow and corresponding memory-side address-generation logic that exploits GEMM locality and enables long-running PIM kernels despite the complex address-mapping functions employed by the CPU that would otherwise destroy locality. Our evaluation of StepStone variants at the channel, device, and within-device PIM levels, along with optimizations that balance parallelism benefits with data-distribution overheads demonstrate $12\times$ better minimum latency than a CPU and $2.8\times$ greater throughput for strict query latency constraints. End-to-end performance analysis of recent recommendation and language models shows that StepStone PIM outperforms a fast CPU (by up to $16\times$) and prior main-memory acceleration approaches (by up to $2.4\times$ compared to the best prior approach).

\end{abstract}

\section{Introduction}
\label{sec:intro}

With the evolution of deep learning (DL), artificial intelligence is being widely used in many internet services. \meadd{We describe a new approach for reducing the latency of such DL inference tasks by accelerating their fully-connected layers with a \emph{processing in/near memory} (PIM) approach.} 
Park et al.~\cite{park2018deep} report that for important personalized recommendation and natural language DL inference workloads, a large fraction of DL-related data-center cycles (42\%) are spent executing fully-connected (FC) layers in Facebook data centers.

FC layers are executed as matrix-matrix multiplication operations (commonly referred to as \emph{GEMM} kernels) and these GEMMs dominate the overall execution time of some workloads~\cite{park2018deep,gupta2020deeprecsys}. GEMMs are commonly considered compute rather than bandwidth bound based on decades of scientific-computing and DL training experience. However, we observe that DL inference GEMMs exhibit two unique traits that leave them memory-bandwidth bound in many cases, and thus amenable to PIM acceleration. 

First, DL inference queries require small-batch execution to meet tight latency constraints, leading to very tall/skinny or short/fat activation matrices. Such matrices offer lower locality, increasing the importance of memory bandwidth. Second, some recommender and language models have billions of parameters (across numerous layers) and it is common for multiple models to be colocated on a single node to improve system efficiency and reduce multi-model query latency~\cite{gupta2020architectural,ke2020recnmp,zhu2019kelp,nishtala2020twig}. As a result, it is common for the larger weight matrices to reside only in main memory, stressing the memory channel when executing on a CPU and often requiring low-bandwidth host-device transfers in systems with accelerators.

Our experiments demonstrate that these GEMM operations are in fact bandwidth-bound on both CPU and GPU systems, and 
describe how they can be accelerated with processing in/near \emph{main} memory (PIM).

We present \emph{StepStone PIM}, which is integrated within the CPU main memory system and solves the dual challenges of utilizing available GEMM locality and sharing data with the CPU under its sophisticated XOR-based DRAM address mapping scheme. Hence, StepStone is an appealing datacenter solution because it: (1) better utilizes bandwidth within the memory system; (2) utilizes locality, enabling high performance and efficiency for datacenter DL inference GEMM operations; (3) does not require additional memory devices or capacity, avoiding the exorbitant cost of additional memory and taking advantage of the already-memory resident matrices; and (4) offloads a low-performance workload from the CPU, freeing additional execution capacity for colocated tasks. 

This unique set of StepStone capabilities is, to the best of our knowledge, not available in any prior PIM architecture and research, including in recent work that targets datacenter DL inference or processing in main memory. While recent work explored PIM-acceleration for datacenter DL inference, it focuses on the embedding layers of DL-inference~\cite{kwon2019tensordimm,ke2020recnmp} rather than on the MLP GEMM operations, which require a different approach for exploiting locality.  Prior work that considers integrating PIM accelerators within main memory either requires costly data replication to avoid the DRAM address mapping challenge~\cite{farmahini2015nda,asghari2016chameleon,alian2018nmp} or does not offer the mechanisms to exploit GEMM locality~\cite{ahn2015pim,kim2017toward,ke2020recnmp,cho2020near}. 

We choose a straight-forward PIM microarchitecture for StepStone that follows recent research trends. Our contributions instead lie with four key innovations. \meadd{The first is the StepStone PIM GEMM parallelization and execution flow that is cognizant of the XOR-based DRAM address mapping that otherwise break GEMM locality. The second contribution accelerates the localization and reduction operations of the execution flow without consuming CPU core resources. The third contribution enables long-running locality-conserving PIM GEMM kernels with the new StepStone memory-side address generation logic. Long-running kernels relieve PIM pressure on the memory command channel, enabling high-performance colocated CPU tasks.}

The fourth contribution is identifying and exploiting a new tradeoff opportunity in balancing the performance benefits of parallelization across fine-grained PIM units (PIMs) within DRAM with the data-transfer overheads of the localization/replication and reduction operations necessary for high parallelization. We explore this tradeoff by evaluating \mbox{channel-,} device-, and bank group-level StepStone PIMs.

To summarize our contributions:
\begin{itemize}
\item We identify and demonstrate that small-batch GEMM operations of DL datacenter inference workloads are bandwidth bound on CPUs and GPUs, and can hence benefit from PIM-acceleration 
  (Section~\ref{sec:motiv}).

\item \meadd{We develop the novel StepStone PIM GEMM execution flow that is cognizant of the complex CPU address mapping, thus exploiting GEMM locality and improving performance by $35-55\%$ over a prior PIM architecture that supports complex address mappings~\cite{cho2020near}.}

\item \meadd{We accelerate the localization and reduction operations of our new GEMM flow at the CPU memory controller to improve performance by up to an additional $40\%$.}

\item \meadd{We design the novel memory-side StepStone address generator that enables long-running GEMM kernels to minimize command-channel usage, which improves PIM performance by $5.5\times$ when the CPU executes concurrent memory-intensive tasks.} 

\item \meadd{We identify a new tradeoff opportunity in determining whether to target channel-, device-, or bank group-level PIMs and show benefits of up to $35\%$ in exploiting it.}

\item We present a detailed StepStone PIM evaluation, including end-to-end performance analysis and conclude that StepStone is an appealing datacenter solution because of its low cost (no additional memory devices or capacity), its potential for lower latency and higher throughput, 
  and its ability to dynamically support the execution of larger-batch and colocated tasks on the CPU.
\end{itemize}

Combining all our innovative mechanisms, StepStone is able to substantially outperform a CPU when executing GEMM operations on matrices with dimensions typical in datacenter DL inference workloads: (1) StepStone offers $12\times$ lower minimum GEMM latency for these matrices; (2) $77\times$ higher throughput under the strictest latency constraints that correspond to batch-1 on the CPU but if the CPU is allowed 20\% additional latency for batch-32 execution, the performance benefit drops to $2.8\times$; 
and (3) up to $16\times$ lower end-to-end DL inference latency compared to measured CPU performance.


\section{Motivation and Challenges}
\label{sec:motiv}


\noindent\textbf{\textit{Bandwidth-bound GEMMs.}}
%
%
Matrix-matrix multiplication (GEMM) is commonly regarded as compute bound. However, we observe that GEMM becomes bandwidth-bound and exhibits low CPU/GPU utilization when both: (1) one of the two input matrices is much larger than the other (e.g., A is large while B is ``tall and skinny'') and (2) the large input matrix is in main memory. While rare in traditional linear algebra applications, DL inference tasks in datacenters often meet both conditions.

First, DL inference queries have tight latency constraints that require small batches \cite{park2018deep}.
The corresponding GEMM operations in fully-connected layers therefore multiply a large weight matrix and a small input matrix. Second, the MLP weights are often only found in main memory because either the total size of the MLP parameters exceeds cache capacity (e.g., in recent language models~\cite{brown2020language,devlin2018bert,radford2019language}) and/or multiple models are colocated on a single node~\cite{gupta2020architectural}.


\begin{table}[b]
\center
\caption{\badd{Common DL-inference GEMM dimensions.}}
\label{tab:matrices}
\tabcolsep=0.14cm
\footnotesize
\begin{tabular}{l|l|l|l|l}
 & Model & Description & Weights & Batch Size \\ 
\hline
\multirow{6}{*}{\begin{tabular}[c]{@{}c@{}}LM\end{tabular}} & 
\multirow{3}{*}{\begin{tabular}[c]{@{}c@{}}BERT\end{tabular}}
 & MLP        & 1024$\times$4096   &  \multirow{6}{*}{\begin{tabular}[c]{@{}c@{}}1-8 \cite{park2018deep} \end{tabular}}   \\ 
 & & MLP         & 4096$\times$1024  &     \\ 
 & & Projection  & 1024$\times$1024   &     \\ 
\cline{2-4}
& \multirow{3}{*}{\begin{tabular}[c]{@{}c@{}}GPT2\end{tabular}}
 & MLP        & 1600$\times$6400   &     \\ 
 & & MLP         & 6400$\times$1600  &     \\ 
 & & Projection  & 1600$\times$1600   &     \\ 
\hline
\multirow{4}{*}{\begin{tabular}[c]{@{}c@{}}RM\end{tabular}} &
\multirow{4}{*}{\begin{tabular}[c]{@{}c@{}}DLRM \\(RM3)\end{tabular}}
 & Bottom MLP & 2560$\times$512  &  \multirow{4}{*}{\begin{tabular}[c]{@{}c@{}}1-256 \cite{park2018deep} \end{tabular}} \\ 
 & & Bottom MLP  & 512$\times$32   &  \\ 
 & & Top MLP     & 512$\times$128  &  \\ 
 & & Top MLP     & 128$\times$1    &  \\ 
\end{tabular}
\end{table}

\begin{figure}[t!]
\centering
        \includegraphics[width=0.46\textwidth]{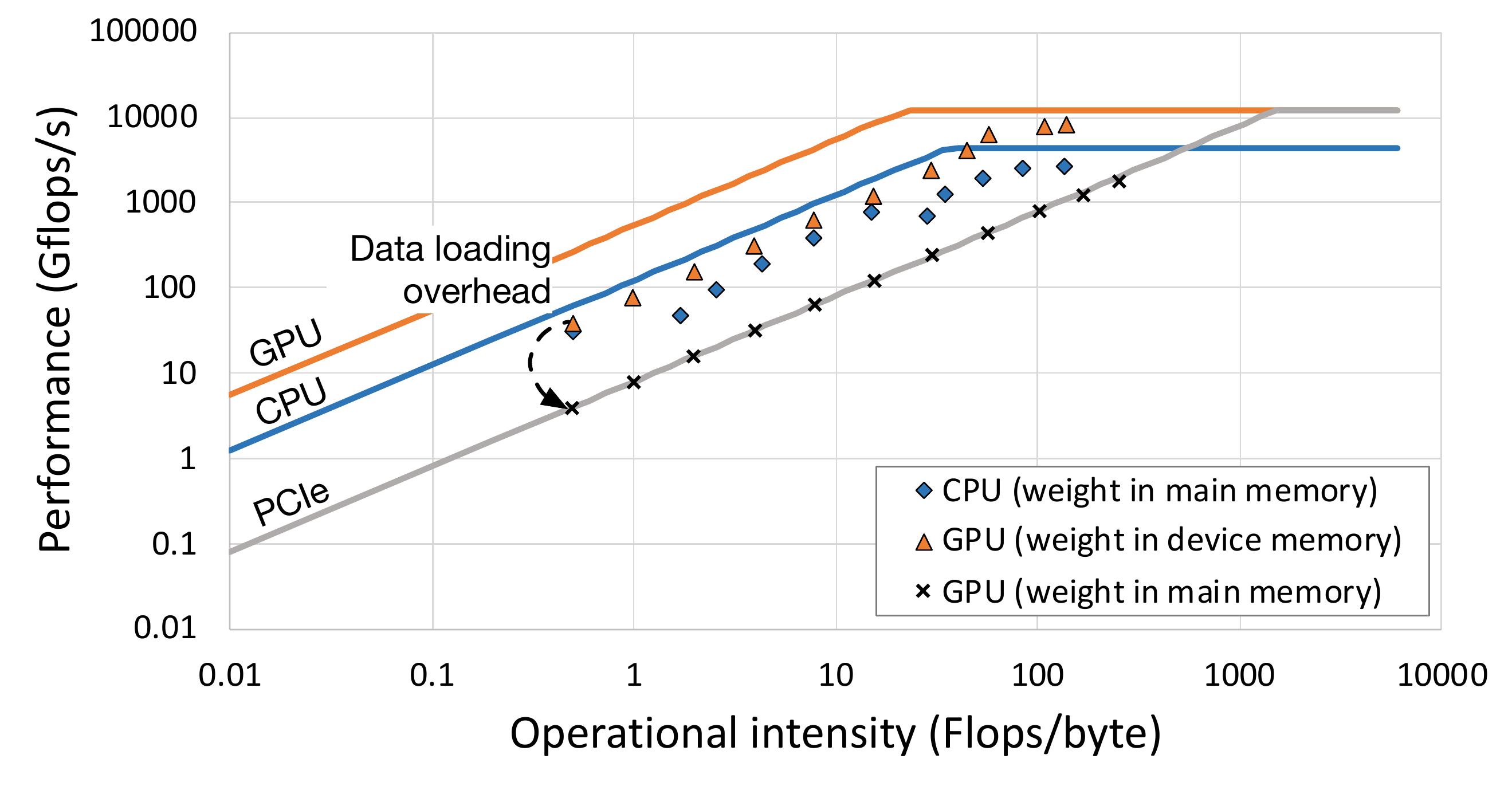}
	\caption{CPU (Intel Xeon Platinum 8280) and GPU (NVIDIA Titan XP) roofline modeling when executing bandwidth-bound GEMM operations of a memory-resident  1024 ${\times}$ 4096 weight matrix with a 4096$\times N$ matrix; $N$ is swept from $1-1024$ in powers of $2$ moving from left to right.}
	\label{fig:roofline}
	\vspace*{-3mm}
\end{figure}

The resulting matrix sizes (Table~\ref{tab:matrices}) are executed inefficiently on CPUs and GPUs as shown by the roofline analysis presented in \fig{fig:roofline}. Each point in the figure corresponds to the performance measured on a 2.7 GHz 28-core Intel Cascade Lake Xeon CPU or an NVIDIA Titan Xp GPU when multiplying a memory-resident 1024$\times$4096 matrix by a cache-resident 4096$\times N$ matrix, where $N$ represents the batch size. The left-most point for each system is when $N=1$ and each point moving right represents a doubling of $N$. We observe that all three systems are bandwidth bound for inference-appropriate batch sizes ($N \lesssim 32$). Further, for such small batches, GPU performance is lower than the CPU if matrix A is in host memory because of the slow host-device bus.

\emph{We conclude that processing in/near memory (PIM) is appealing for these GEMM operations of datacenter DL-inference workloads.}

\begin{figure*}[t!]
\centering
        \includegraphics[width=0.85\textwidth]{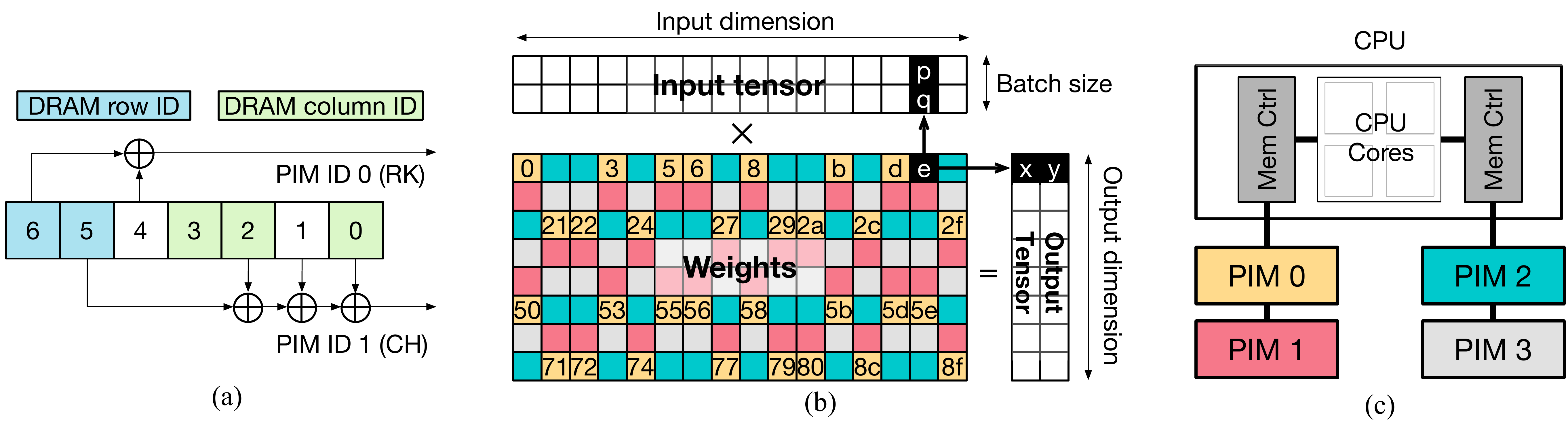}
	\caption{An example of bandwidth-bound GEMM operation with PIM and a toy XOR-based address mapping: (a) toy XOR-based physical-to-DRAM address mapping where addresses refer to contiguous row-major matrix elements; (b) layout of an 8 ${\times}$ 16 matrix with colors indicating element$\rightarrow$PIM unit mapping; (c) example system with rank-level PIMs.}
	\label{fig:example_xor_mapping}
	\vspace*{-3mm}
\end{figure*}

%

\medskip
\noindent\textbf{PIM GEMMs with XOR-based DRAM address mapping.}
We target systems in which main memory is PIM enabled, implying a shared DRAM address space with the CPU. The CPU relies on sophisticated XOR-based DRAM address mappings to exploit bank and channel parallelism by distributing consecutive cache blocks in the physical address space (PA) across DRAM banks and channels. As a result, matrices that are contiguous in the application virtual space are distributed across PIM units (PIMs) in complex patterns. Effective GEMM execution requires exploiting locality and reuse in matrix blocks, which are challenging to identify.

\fig{fig:example_xor_mapping} illustrates this challenge for the toy address mapping of \fig{fig:example_xor_mapping}{a} targeting a system with 4 PIM units (one per rank). Addresses refer to elements of the large matrix shown in \fig{fig:example_xor_mapping}{b}, which is laid out row major in contiguous memory addresses. Logical blocks of the matrix do not form blocks within each PIM. For example, element \texttt{0e} of the weight matrix (marked in black) is mapped to PIM0 and is multiplied by elements \texttt{p} and \texttt{q} from the input tensor to modify elements \texttt{x} and \texttt{y} of the output tensor. These same elements of the input tensor are also needed when reading element \texttt{5e} of the weight matrix and the same two output-tensor elements when reading weights \texttt{00}, \texttt{03}, \texttt{05}, \texttt{06}, \texttt{08}, \texttt{0b}, and \texttt{0d}. Utilizing this locality requires the PIMs to correctly map between contiguous DRAM addresses within each PIM and the corresponding addresses of the input and output tensors.


\meadd{Prior approaches to this challenge fall into one of three categories. The first avoids the challenge altogether by maintaining a copy of the data that is stored in a PIM-friendly layout and not accessed by the CPU~\cite{he2020newton,kim2016neurocube,kwon2019tensordimm}. This either duplicates substantial data arrays (possibly $>100$GiB)~\cite{brown2020language,radford2019language,DLRM19} or prevents the CPU from assisting with requests that can tolerate higher response latency~\cite{gupta2020deeprecsys}. Furthermore, a different layout is needed for channel-, device-, and bank group-level PIMs. This either forces even more replicas or prohibits optimizations that dynamically choose the PIM level based on input characteristics (e.g., as in the XLM language model~\cite{lample2019cross}).}

The second approach requires the CPU to transfer this correspondence information to the PIMs for each cache block the PIM processes~\cite{ahn2015pim,kim2017toward}. The CPU sends special DRAM command packets that include operand and opcode information to each PIM unit and all the transactions related to PIM execution are controlled by the host. PIMs are isolated from the address mapping concerns, but performance scalability is poor because: (1) channel bandwidth for sending PIM command packets saturates rapidly, (2) CPU resources are required for address generation, and (3) the frequent PIM command packets severely interfere with CPU memory traffic~\cite{cho2020near}. 

  The third approach, proposed by Cho et al.~\cite{cho2020near}, aligns long vector PIM operands in memory such that all kernel operands follow the same interleaving pattern after the XOR address mapping. In this way, both the CPU and the vector-oriented PIM can process the same data. However, this vector-oriented approach cannot exploit the GEMM kernel locality. Vector-oriented execution splits the GEMM into multiple matrix-vector (GEMV) operations, requiring a larger number of kernel invocations. The straightforward implementation also requires copies across PIM units to ensure all data is local. The standalone (non main-memory) Newton PIM accelerator~\cite{he2020newton} also follows this approach. We observe that a different execution flow can be used to block both the input and output matrices to reduce copy overhead. We explain our StepStone PIM GEMM in the following section.

\section{StepStone PIM}
\label{sec:gemm}

\textit{StepStone PIM} enables independent GEMM execution with PIMs under any XOR-based memory-system address mapping. In StepStone PIM, the weight matrix is partitioned and assigned to PIMs based on the underlying address mapping, maintaining internal contiguity and enabling temporal locality when each PIM unit works on its GEMM blocks. From the CPU perspective, the PIMs appear to skip within the address space and only step on those ``stones'' (i.e. cache blocks) that are physically local to them.

\subsection{StepStone Architecture}
\label{subsec:baseline_system}
\begin{figure}[t!]
	\subfloat[Baseline PIM system.]{%
  		\includegraphics[width=\columnwidth]{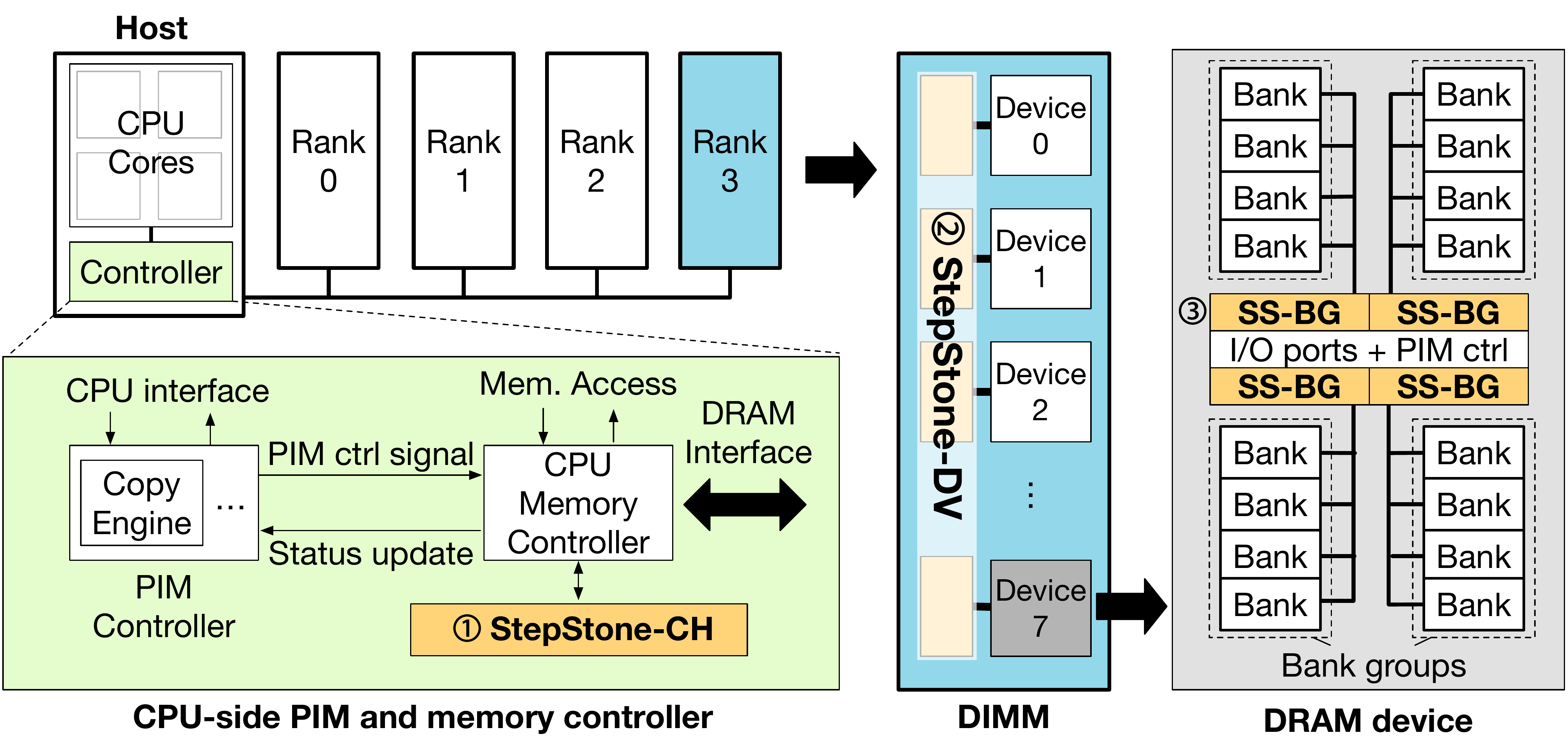}%
	}
	
	\centering
	\subfloat[StepStone PIM architecture.]{%
  		\includegraphics[width=0.7\columnwidth]{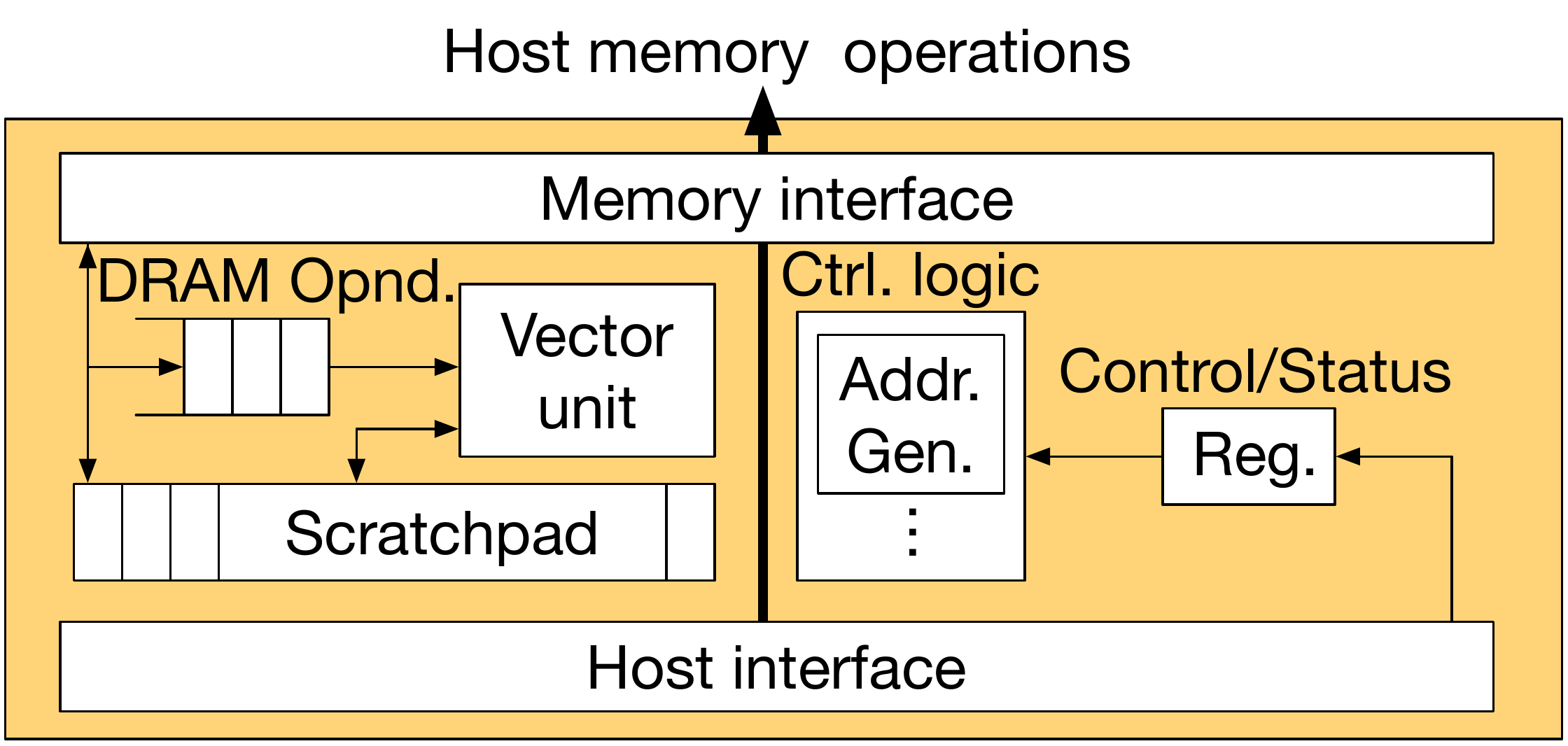}%
	}
	\caption{Overview of the StepStone PIM System.}
	\label{fig:baseline_system}
	\vspace*{-3mm}
\end{figure}

The StepStone PIM architecture is, for the most part, a standard PIM. The innovation lies in how we map GEMM operations and in our unique address-generation algorithm, both discussed later in this section. StepStone is comprised of a host-side PIM controller that interfaces with PIM units (PIMs) through the memory channel to control their operation using memory-mapped PIM-side registers. As shown in \fig{fig:baseline_system}{a}, PIM units (PIMs) can be integrated with: (1) DRAM itself, e.g., at the bank-group level (StepStone-BG); (2) with a memory module, e.g., associated with each DRAM device or buffer chip (StepStone-DV);\footnote{\meadd{This is a cheap rank-level PIM with no inter-device communication.}} and/or (3) with a memory channel controller (StepStone-CH). We consider all three integration levels. Note that StepStone-BG accounts for device-level timing parameters such as tRCD and tFAW using control logic at the I/O port of each device.

Each PIM unit (\fig{fig:baseline_system}{b}) includes SIMD/vector units, scratchpad memory, interfaces, control logic to execute the GEMM kernel command (sub-GEMM to be more precise), and a local-address generation unit. The pipeline is sufficiently deep to hide address generation and access latencies (20 stages in our case). When $N$ \meadd{(e.g., the batch dimension)} is large, performance is bound by the SIMD width. While wide SIMD units are desirable, arithmetic performance must be balanced with area and power constraints.

Following prior work, we aim for $0.15\mathit{mm}^2$ for each StepStone-BG unit~\cite{gu2020ipim} and $1.2\mathit{mm}^2$ for StepStone-DV~\cite{asghari2016chameleon}. We aim for a $2:1$ ratio between SIMD area and scratchpad area and assume additional logic is of comparable size to the scratchpad. We estimate functional unit and scratchpad area and power at the device-level with the values reported for iPIM~\cite{gu2020ipim} and at the device and channel levels following the methodology of Lym et al.~\cite{lym2019mbs}. This analysis yields nominal values of 8-wide SIMD with $8$KB scratchpad capacity for each StepStone-BG unit (4 PIMs per DRAM device), and 32-wide SIMD with $32$KB scratchpad capacity per StepStone-DV PIM unit. For StepStone-CH, we keep the same bandwidth to arithmetic performance ratio as StepStone-DV: 256-wide SIMD units and $256$KB scratchpad capacity. We consider all three cases and conclude that StepStone-CH offers the lowest performance and requires the largest die area. 

One other component is the replication/reduction unit within the PIM controller, which is used to accelerate the distribution of matrix B and reduction of partial values of C, which are required for the GEMM execution described below.


\subsection{StepStone GEMM Execution}
\label{subsec:gemm_overview}

\begin{figure*}[t!]
\centering
        \includegraphics[width=0.8\textwidth]{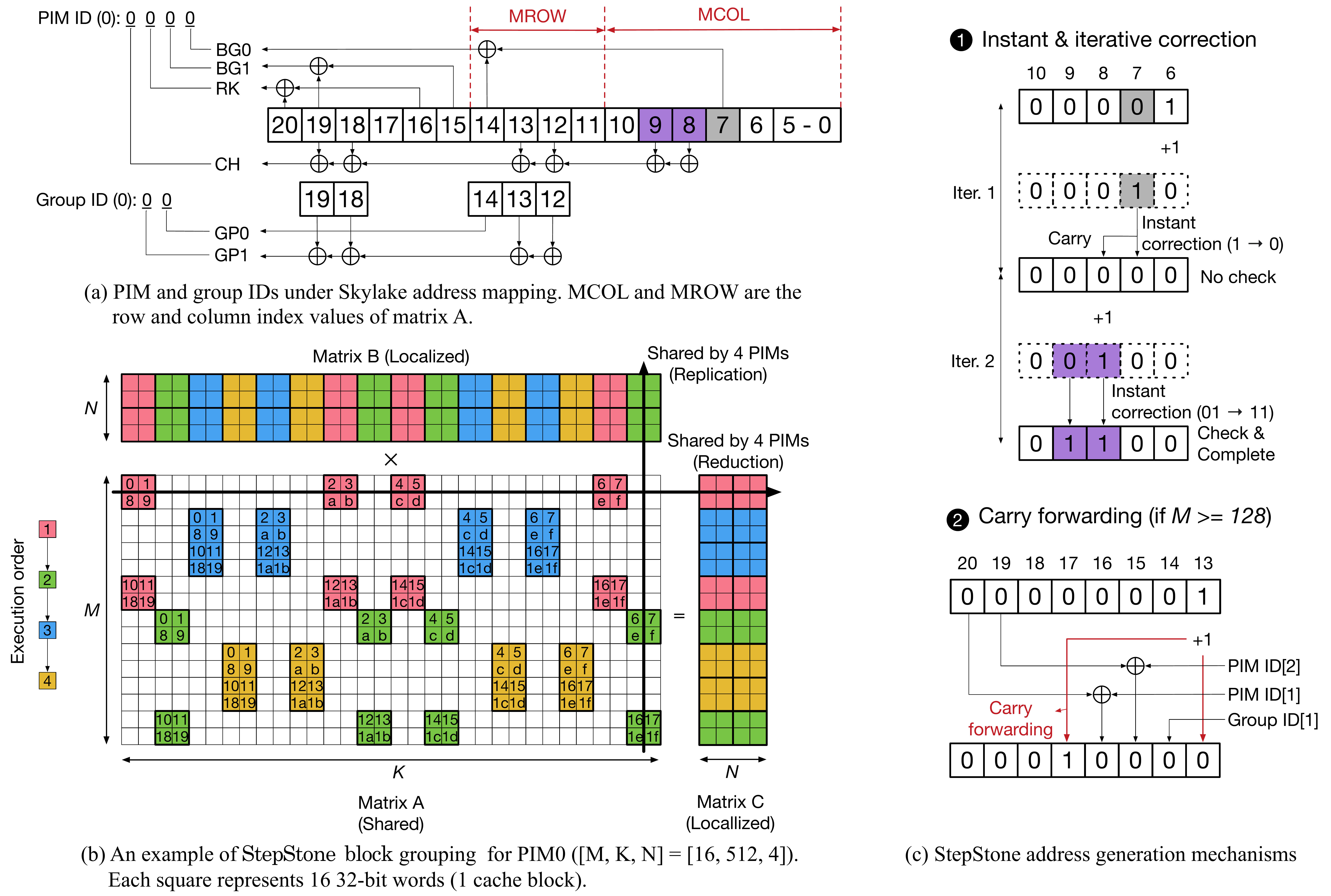}
	\caption{Overview of GEMM execution with StepStone PIM.}
	\label{fig:overview}
	\vspace*{-3mm}
\end{figure*}

GEMM execution starts with the large weight matrix A stored contiguously in the virtual and physical address spaces in row-major order.\footnote{We assume that the matrix dimensions are powers of two; matrices with non-power-of-two dimensions are either padded or execution is partitioned/serialized into smaller, power-of-two matrices.}
Therefore, A is distributed across memory devices based on the DRAM address mapping as shown in \fig{fig:overview} (for the Intel Skylake address mapping~\cite{pessl2016drama} on StepStone-BG and depicting only the elements of A that map to PIM0, which we refer to as partition A0). A0 must be multiplied with elements of B and accumulated into elements of C. To maximize parallelism, we first localize private copies of B and C into each PIM unit, also shown in the figure. Localizing data copies the data into a pre-allocated per PIM-unit memory region. Execution then proceeds with a partial dot product across rows of A0 with columns of B (B is shown transposed in the figure).

However, recall the dual challenges of identifying the corresponding indices in B and C as A0 is advanced while maximizing reuse during execution. We address these challenges grouping together cacheline-sized memory blocks (``cache blocks'') into \emph{block groups} that follow the same DRAM address mapping distribution. We note that the grouping depends both on the address mapping and the matrix dimension. Each block group is shaded a different color in \fig{fig:overview}{b}. 


\label{subsec:gemm_grouping}
\medskip
\noindent\textbf{StepStone locality.}
To maximize reuse, each element of B should be multiplied with as many elements of A before overwritten in the buffer. We achieve this by executing one block group at a time: cache blocks within each group across rows reuse elements of C while those along columns reuse elements of B. No reuse exists between groups. 

The number of groups required to maximize locality is determined by the number of PIM ID bits that are impacted by addresses within the matrix. For example, the matrix in \fig{fig:overview} is 16$\times$512 4B words and starts at physical address 0, thus locations within this matrix span the lower 15 address bits. Within these bits, bits 7 and 14 determine one bank-group bit (BG0, which is also PIM ID bit 0) and bits 8, 9, 12, and 13 affect the channel bit (PIM ID3). The other PIM ID bits are fixed for all locations within this matrix. We further note that a group spans entire rows to maximize locality. We therefore exclude address bits associated with matrix columns (MCOL) from defining the group ID bits (GP0 and GP1 in the figure).

\label{sec:impl}

\begin{figure}[t!]
\centering
        \includegraphics[width=0.48\textwidth]{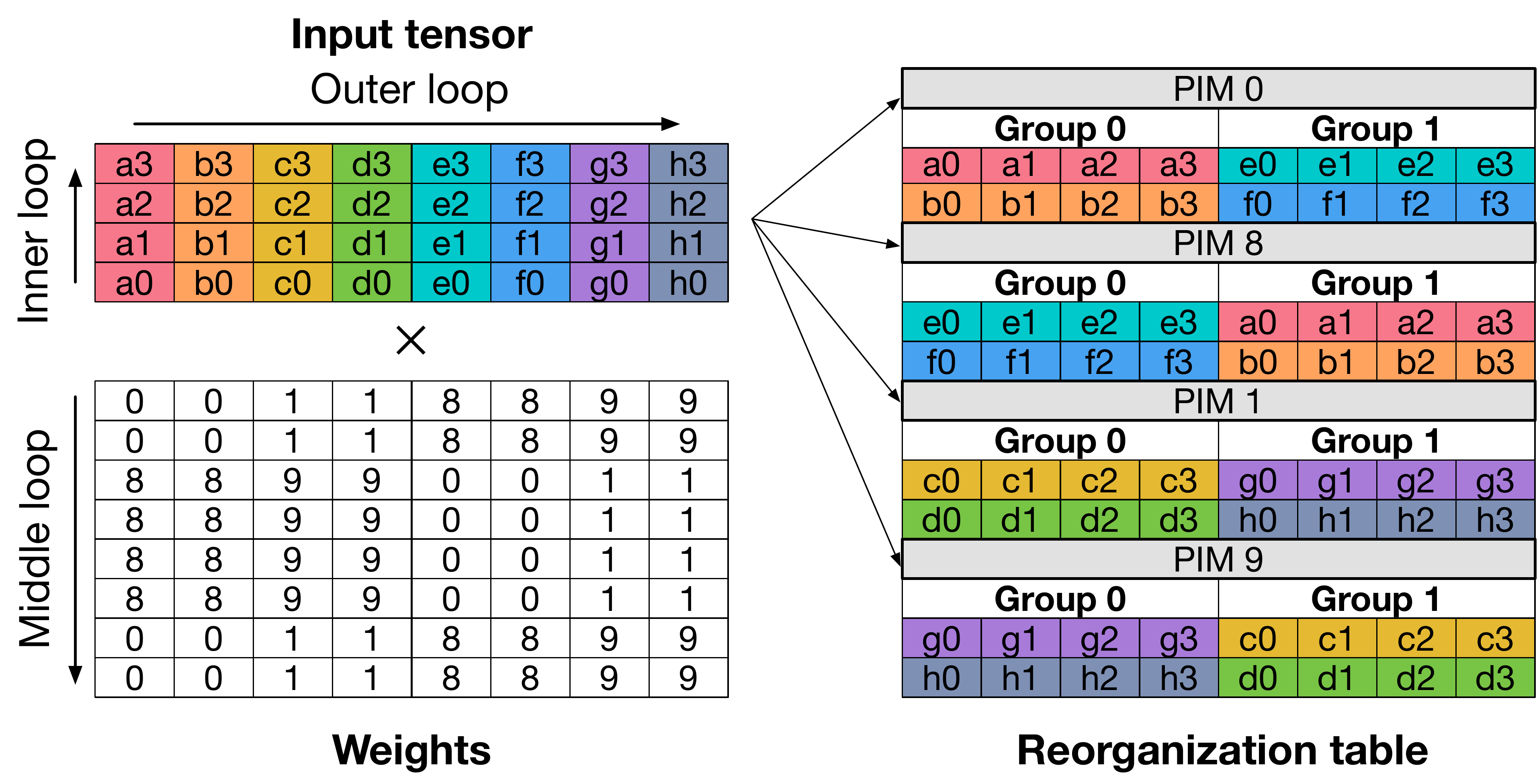}
	\caption{Input-matrix Reorganization.}
	\label{fig:replication}
	\vspace*{-3mm}
\end{figure}

\medskip
\noindent\textbf{Localizing matrices B and C.}
\label{subsec:mech_access_pvt_mem}
Matrix B is (partially) replicated and localized to the different PIMs before execution begins, and the localized partial-result C matrices are reduced after the GEMM. The replication and reduction, along with data reorganization for spatial locality within each PIM unit are handled by the host and accelerated with a simple DMA engine at the PIM controller.
The operation of this engine is illustrated in \fig{fig:replication} for localizing matrix B for a portion of matrix A that is distributed across PIMs 0, 1, 8, and 9. Matrix B is again represented transposed in the figure and consecutive elements in each of its rows appear as columns (e.g., a0 - a3).

During localization, the engine reorganizes the input matrix for each PIM unit such that accesses are sequential during its group-based execution. The outer-most loop iterates over columns of A, localizing the rows of B (appear as columns in B$^T$) needed for each column in the PIMs and block groups it maps to. The PIM and group IDs are computed based on the mappings illustrated in \fig{fig:overview}{a}. Each cache block of B is read once and then copied to all its relevant PIM-local addresses. Reductions follow a similar execution flow.

\subsection{Overall Execution Flow of StepStone GEMM}
\label{subsec:mech_pim_flow}
The execution flow of complete GEMM is shown in Algorithm~\ref{algo:gemm}. After localization, the input matrices are all aligned in the DRAM accessible by each PIM unit and execution proceeds in group order. A slight complication arises when A is very large such that not all elements of B and C that correspond to a group can be buffered within the scratchpad. In such cases, to still utilize locality, we further block each group. This blocking can be across rows, maximizing reuse of C, and/or across columns. We process blocks of rows first 
because C offers greater reuse as it is both read and written.

\meadd{The inner-most GEMM call is coarse-grained for the full StepStone PIM with the mapping-aware AGEN unit, but is split into multiple dot product operations without this innovative AGEN logic. In this way we can isolate  the contributions of our algorithm mapping and hardware mechanism when evaluating StepStone PIM compared to prior PIM architectures, like Chopim~\cite{cho2020near} (we denote our StepStone GEMM flow on Chopim as \emph{enhanced Chopim} or eCHO).} 

Note that address generation with partitioning is slightly different than as described for unpartitioned groups execution. When crossing different column partitions (groups of columns that partition a row into multiple blocks), address generation must skip over those columns belonging to different partitions. This is simple to do and only requires modifying the address-generation rules to account for group and partition ID. 

\subsection{StepStone Address Generation}
\label{subsec:gemm_addr_gen}
Within a single cache block, the address is a simple increment, but once the value of a bit that determines the PIM ID is modified, that contiguous physical address must be skipped until the next physical address that maps to the same PIM unit and block group is reached. A simple iterative approach of incrementing the address until the address is again within this same block and PIM ID, the number of iterations required when the number of PIMs is large introduces bubbles into the execution pipeline and degrades performance.\footnote{We assume that the CPU address mapping is available for PIMs either by reverse engineering, by CPU vendors building the PIMs, or by agreement.}

\meadd{We propose new  \textit{increment-correct-and-check} AGEN logic that skips to the next closest address with the same PIM and group IDs (after the simple increment falls outside the target IDs). We do this by ensuring that those address bits that are XORed to determine each ID bit always maintain their parity. We can thus skip incrementing bits that are lower than the lowest ID-affecting address bit. The AGEN logic iterates over the ID-affecting bits (from LSB to MSB), each time incrementing the next ID-affecting bit and checking whether the PIM and group IDs match their target values.}

\meadd{The number of iterations is limited to the number of ID-affecting bits, but can be further reduced with two additional rules. The first rule applies for adjacent address bits that both affect the same ID bit. When the lower of the two is incremented, the upper must be as well to maintain parity. This can be done directly, saving one iteration. The second rule applies for chains of contiguous address bits that each affect a different ID bit. In this case, when the first is incremented, the carry it propagates will have to corrected in multiple iterations to maintain the parity corresponding to each bit in the chain. Thus, the chain can be skipped with the carry simply directly propagated to the next-higher address bit. These rules are shown in}
\fig{fig:overview}{c}. The top part of the figure illustrates the first rule of instantly correcting from \texttt{01} to \texttt{11}, and the bottom part illustrates the second rule of forwarding the carry from bit 13 to 17 since bit 14-16 affect different ID bits.

\begin{algorithm}[h]
  \SetAlgoLined
  localize(B)\;
  localize(C)\;
 \For{rpart in row\_partitions}{
  buffer\_fill(C)\;
  \For{grp in block\_groups}{
  	\For{cpart in col\_partitions}{
  	  buffer\_fill(B)\;
	  \If{StepStone}{
  	  	StepStone\_GEMM\;
  	  }
  	  \ElseIf{eCHO}{
  	  	\For{row in cpart}{
  	  			DOT(row)\;
  	  	}
  	  }
  	}
  }
  buffer\_drain(C)\;
}
reduce(C)\;
 \caption{Group-based GEMM with partitioning.}
 \label{algo:gemm}
\end{algorithm}

\subsection{Optimizations}
Multiple optimizations over the basic flow described above are possible, including fusing multiple kernel executions for matrices that are not powers of two, balancing parallel execution with the overheads of localization and reduction, and choosing a PIM level for execution (StepStone-BG vs.\ StepStone-DV). For brevity, we only discuss the latter two.

\medskip
\noindent\textbf{Choosing the PIM level.}
Bank-group level StepStone-BG offers the highest potential performance when the GEMM is severely bandwidth bound (very small batches) because it accesses underutilized bandwidth within a DRAM device. An interesting tradeoff emerges as bandwidth constraints decrease with somewhat larger batches. The arithmetic intensity (data resuse per operation) in StepStone scales with the batch size ($N$) up to the SIMD width of each PIM unit. This results in comparable arithmetic execution times for $1\leq N \leq 16$ in StepStone-BG and for  $1\leq N \leq 32$ in StepStone-DV (though obviously the execution times differ between the two PIM levels). At the same time, the overheads of localization and reduction increase with $N$ and with the number of PIMs (number of block groups).

An optimization opportunity emerges for choosing between bank-group and device level PIMs as a result. The best PIM level depends on the matrix dimensions and the address mapping as these determine the number of block groups, and hence the localization and reduction overheads. We demonstrate this tradeoff in Section~\ref{sec:eval} and generally find that StepStone-BG is best when $N\leq 16$. Note that we do not discuss the algorithm for choosing the PIM level, but note that a simple heuristic that estimates execution times and overheads based on available bandwidth and transferred data volumes works well.

\medskip
\noindent\textbf{Small weight matrices.}
\label{subsec:opt_tradeoff}
A similar tradeoff between arithmetic performance and localization and reduction overhead exists when the matrices are relatively small. In such cases, it is preferable to only use a subset of the available PIMs, trading off lower arithmetic performance for reduced overheads.  We show that this optimization can yield a $\sim 25\%$ performance improvement in some cases (Section~\ref{subsec:eval_tradeoff}). Another optimization for relatively small matrix A is that CPU can directly write to and read from PIM's scratchpad memory when the matrix B and C fits in it. This reduces the time to move data between DRAM and the scratchpad memory. 

\label{subsec:mech_malloc}
Optimizing execution to only utilize a subset of the PIMs comes with additional considerations when allocating memory for the large distributed input matrix A. Specifically, A must remain contiguous in virtual memory yet be mapped to just a subset of the PEs. Because the address mapping and size of the matrix is known, it is possible to allocated physical memory frames to satisfy this mapping constraint as long as the PIM ID bits that are ignored (for subsetting) are not affected by virtual-address offset bits. In other words, this is possible with base pages only (e.g., 4KB pages).
Enforcing the mapping constraints to maintain alignment with the PIMs can be done using the proposed coloring interface introduced by Cho et al.~\cite{cho2020near} and by modifying the application's memory allocator.

For example, if the goal is to execute on half the PIMs of StepStone-BG with the Skylake mapping, we keep BG0 fixed for the entire physical allocation of A. This is achieved by allocating virtual memory at 32KB granularity rather than the minimum 4KB granularity. Additionally, we must ensure that contiguous virtual addresses remain aligned in the DRAM space and therefore must also ensure that the other PIM ID bits follow a consistent mapping. We do that by coloring those bits in a way that the OS-level frame allocator maintains alignment, as proposed for Chopim~\cite{cho2020near}.


\section{Methodology}
\label{sec:method}

\noindent\textbf{System configuration.}
Table \ref{tab:eval_config} summarizes our system configuration and DRAM timing parameters. Our DRAM model is based on the DDR4 model of the Ramulator detailed DRAM simulator~\cite{kim2016ramulator}. We implement StepStone-CH, StepStone-DV, and StepStone-BG PIMs inside Ramulator with detailed pipeline and timing models. We emulate our memory allocator and add an address translation engine into the PIM controller on the CPU (\ref{subsec:baseline_system}); address translation is infrequent (once per coarse-grained PIM command) because contiguous physical regions are allocated for PIM execution. To validate our execution flow, we modify Ramulator to read from and write values to memory and check the final output against pre-calculated results. We also compare all addresses from the StepStone AGEN logic with a pre-generated address trace for each PIM. For all the GEMM operations with StepStone PIM, we assume the input activations reside in the CPU caches and are therefore localized to the active PIMs. In the same way, we assume the final GEMM results are reduced by the CPU. 

We use the XOR-based address mappings described in DRAMA~\cite{pessl2016drama}, acquired by reverse-engineering Intel and Samsung CPUs. To show the impact of address mapping on the same DDR4 model, we modify the address mapping of Exynos, Dual-socket Haswell, Ivy Bridge, and Sandy Bridge based on the randomized method (PAE) proposed by Liu et al.~\cite{liu2018get}. By default we we use Skylake's address mapping. To measure GEMM performance on real machines, we use an Intel Xeon Platinum 8280 (CPU) with Intel oneDNN~\cite{onednn} and an NVIDIA TitanXP (GPU) with CUTLASS~\cite{cutlass}. 

The area and power of the SIMD units are estimated based on Lym et al.~\cite{lym2019mbs} for StepStone-DV and StepStone-CH and iPIM~\cite{gu2020ipim} for the SIMD units of StepStone-BG. We use Cacti 6.5~\cite{muralimanohar2009cacti} to estimate the area and power of scratchpad memory.

\noindent\textbf{Workloads.}
We choose matrix sizes and aspect ratios to clearly show their impact on performance. By default, we use 1024 ${\times}$ 4096. For end-to-end performance evaluation, we use the 4 different DL models summarized in Table~\ref{tab:eval_config}. Since our goal is to accelerate GEMMs with small batch sizes, we vary the batch size from 1 to 32. We refer to input and output activations as matrix B and C, respectively, and we refer to the largest matrix as weight matrix A.

\meadd{
  We demonstrate the benefits of long-running kernels for concurrent CPU and PIM execution in a colocation scenario of the default 1024 ${\times}$ 4096 GEMM with a mix of the mcf, lbm, omnetpp, and gemsFDTD from SPEC CPU 2017~\cite{panda2018wait}. The CPU applications are modeled with gem5~\cite{binkert2011gem5} (4 OOO x86 4GHz cores with fetch/issue width of 8, a 64-deep LSQ, and a 224-entry ROB), similarly to \cite{cho2020near}.}

\noindent\textbf{Comparisons.}
We compare our approach with two existing PIM approaches, PEI~\cite{ahn2015pim} and Chopim~\cite{cho2020near}, which are capable of accelerating GEMMs in main memory and leveraging multi-level PIMs with one data layout. To make a fair comparison, we use our baseline PIM system (\fig{fig:baseline_system}) for all approaches and only vary localization/reduction mechanisms and PIM kernel granularity. For PEI, each PIM instruction is used to process one cache block and all the other operands needed for executing the PIM instruction is written to scratchpad memory by the CPU. \meadd{We evaluate two versions of Chopim. The baseline naive Chopim (nCHO) follows the GEMV mapping approach (Section~\ref{sec:motiv}). We also use our newly-developed StepStone flow with an ``enahnced'' Chopim (eCHO). This eCHO configuration exploits locality as well as StepStone PIM, but requires more frequent kernel calls (Algorithm~\ref{algo:gemm}) and does not use accelerated localization and reduction operations. }

\begin{table}[!t]
	\centering
	\caption{Evaluation parameters.}
	\noindent\resizebox{\linewidth}{!}{
		\tabulinesep=0.6mm
		\begin{tabu}{c|c|c|c|c}
			\hline
			\arrayrulecolor{white}\hline
			\arrayrulecolor{white}\hline
			\arrayrulecolor{white}\hline
			\arrayrulecolor{black}\hline
			\rowfont{\normalsize}
			\multicolumn{5}{c}{PIM configurations} \tabularnewline
			\hline
      		StepStone-BG & \multicolumn{4}{c}{\makecell{8-width SIMD, 8KB scratchpad (per DRAM device), 1.2GHz}} \tabularnewline
      		StepStone-DV & \multicolumn{4}{c}{\makecell{32-width SIMD, 32KB scratchpad (per buffer chip), 1.2GHz}} \tabularnewline
      		StepStone-CH & \multicolumn{4}{c}{\makecell{256-width SIMD, 256KB scratchpad (per channel), 1.2GHz}} \tabularnewline
			\hline 
			\arrayrulecolor{white}\hline
			\arrayrulecolor{white}\hline
			\arrayrulecolor{white}\hline
			\arrayrulecolor{black}\hline
			\rowfont{\normalsize}
            \multicolumn{5}{c}{Address mappings} \tabularnewline
			\hline
      		ID = 0 & \multicolumn{4}{c}{\makecell{Exynos-like address mapping (modified)}} \tabularnewline
			\hline 
			1 & \multicolumn{4}{c}{\makecell{Haswell-like address mapping (modified)}} \tabularnewline
			\hline 
			2 & \multicolumn{4}{c}{\makecell{Ivy Bridge-like address mapping (modified)}} \tabularnewline
			\hline 
			3 & \multicolumn{4}{c}{\makecell{Sandy Bridge-like address mapping (modified)}} \tabularnewline
			\hline
			4 & \multicolumn{4}{c}{\makecell{Skylake address mapping (baseline)}} \tabularnewline
			\hline 
			\arrayrulecolor{white}\hline
			\arrayrulecolor{white}\hline
			\arrayrulecolor{white}\hline
			\arrayrulecolor{black}\hline
			\rowfont{\normalsize}
            \multicolumn{5}{c}{DRAM timing parameters (DDR4-2400R, 4GB, x8)} \tabularnewline
			\hline

			\multicolumn{5}{c}{\makecell{tBL=4, tCCDS=4, tCCDL=6, tRTRS=2, tCL=16, tRCD=16,\\
			tRP=16, tCWL=12, tRAS=39, tRC=55, tRTP=9, tWTRS=3,\\
      tWTRL=9, tWR=18, tRRDS=4, tRRDL=6, tFAW=26}} \tabularnewline
			\hline 
			\arrayrulecolor{white}\hline
			\arrayrulecolor{white}\hline
			\arrayrulecolor{white}\hline
			\arrayrulecolor{black}\hline
			\rowfont{\normalsize}
			\multicolumn{5}{c}{Energy components} \tabularnewline
			\hline
			\multicolumn{5}{c}{\makecell{In-device RD/WR (11.3pJ/b), Off-chip RD/WR(25.7pJ/b)s, \\
			CH/DV/BG SIMD (11.3,11.3,11.3nJ/op), \\
			CH/DV/BG Scratchpad (0.03/0.1/0.3 nJ/access)}} \tabularnewline
			\hline 
			\arrayrulecolor{white}\hline
			\arrayrulecolor{white}\hline
			\arrayrulecolor{white}\hline
			\arrayrulecolor{black}\hline
			\rowfont{\normalsize}
			\multicolumn{5}{c}{\badd{DL inference parameters}} \tabularnewline
			\hline
			\multirow{1}{*}{\begin{tabular}[c]{@{}c@{}}RM\end{tabular}}
			& DLRM \cite{DLRM19}  & \multicolumn{3}{c}{\makecell{RM3, Bottom MLP (2560-512-32), \\ Top MLP (512-128-1), bsz=4}} \tabularnewline
			\hline 
			\multirow{3}{*}{\begin{tabular}[c]{@{}c@{}}LM\end{tabular}}
			& BERT \cite{devlin2018bert} & \multicolumn{3}{c}{\makecell{Text classification (WNLI), 24 transformer blocks, \\MLP (1024-4096-1024), seq. length=8, bsz=4\\\#attention heads=16}} \tabularnewline
			\cline{2-5}
			& GPT2 \cite{radford2019language}  & \multicolumn{3}{c}{\makecell{Text generation, 48 transformer blocks, \\MLP (1600-6400-1600), seq. length=8, bsz=4}} \tabularnewline
			\cline{2-5}
			& XLM \cite{lample2019cross} & \multicolumn{3}{c}{\makecell{Text generation, 12 transformer blocks, \\MLP (2048-8192-2048), seq. length=8, bsz=4}} \tabularnewline
			\hline
			\arrayrulecolor{white}\hline
			\arrayrulecolor{white}\hline
			\arrayrulecolor{white}\hline
			\arrayrulecolor{black}\hline
		\end{tabu}
	}
	\label{tab:eval_config} 
	\vspace*{-4mm}
\end{table}


\section{Evaluation Results}
\label{sec:eval}

In this section, we demonstrate the throughput and latency benefits of StepStone over either a CPU or GPU, evaluate the impact of address mapping and scratchpad capacity, and analyze the tradeoff between arithmetic performance and overheads as the number of active PIM units (PIMs) is varied.

\subsection{StepStone PIM Performance Benefits}
\label{subsec:eval_latency}
\begin{figure}[t!]
\centering
        \includegraphics[width=0.48\textwidth]{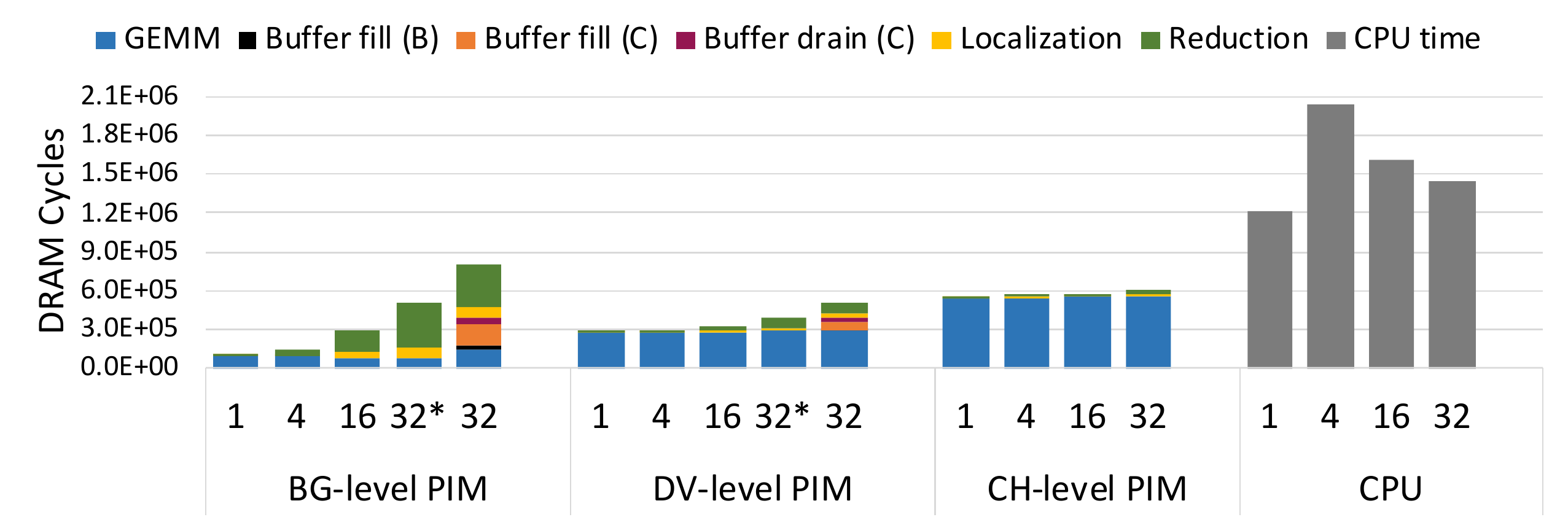}
	\caption{GEMM Latency comparison between different PIM options of StepStone PIM and the CPU. The configurations with relaxed area constraints are labeled with * (i.e. enough ALUs and large enough scratchpad memory).}
	\label{fig:eval_latency}
	\vspace*{-3mm}
\end{figure}

We first compare the performance of StepStone PIM to a 2.7GHz 28-core Intel Xeon Platinum 8280 Cascade Lake CPU with a representative 1024 $\times$ 4096 weight matrix (\fig{fig:eval_latency}). StepStone PIM offers tremendous benefits for latency and throughput targets. When considering the minimum latency batch-1 execution, StepStone-BG with a PIM unit per bank group offers far superior latency: $2.8\times$ better than the device-level StepStone-DV and $12\times$ better than the CPU.

Alternatively, we consider maximum throughput under a latency constraint. When the latency constraint is set to the minimal latency of the CPU (CPU with batch-1), StepStone-DV offers $77\times$ higher throughput ($32\times$ more samples at about $40\%$ less time). If we allow a larger-area PIM with larger scratchpads, performance is improved even further to $96\times$. If we relax the latency constraint and allow the CPU $1.2\times$ more time to complete an inference task, which allows batch-32 on the CPU, the performance benefit of StepStone-DV drops to $3\times$ ($3.5\times$ with a larger scratchpad).
While we use the highly-optimized Intel OneDNN library on the CPU, the performance we observe falls short of the channel-level StepStone-CH, which can fully utilize the memory-system bandwidth. Still, the finer-grained StepStone-DV (which can be implemented in buffer chips) offers substantially better performance and StepStone-BG offers far lower minimum latency. 


\begin{figure}[t!]
\centering
        \includegraphics[width=0.48\textwidth]{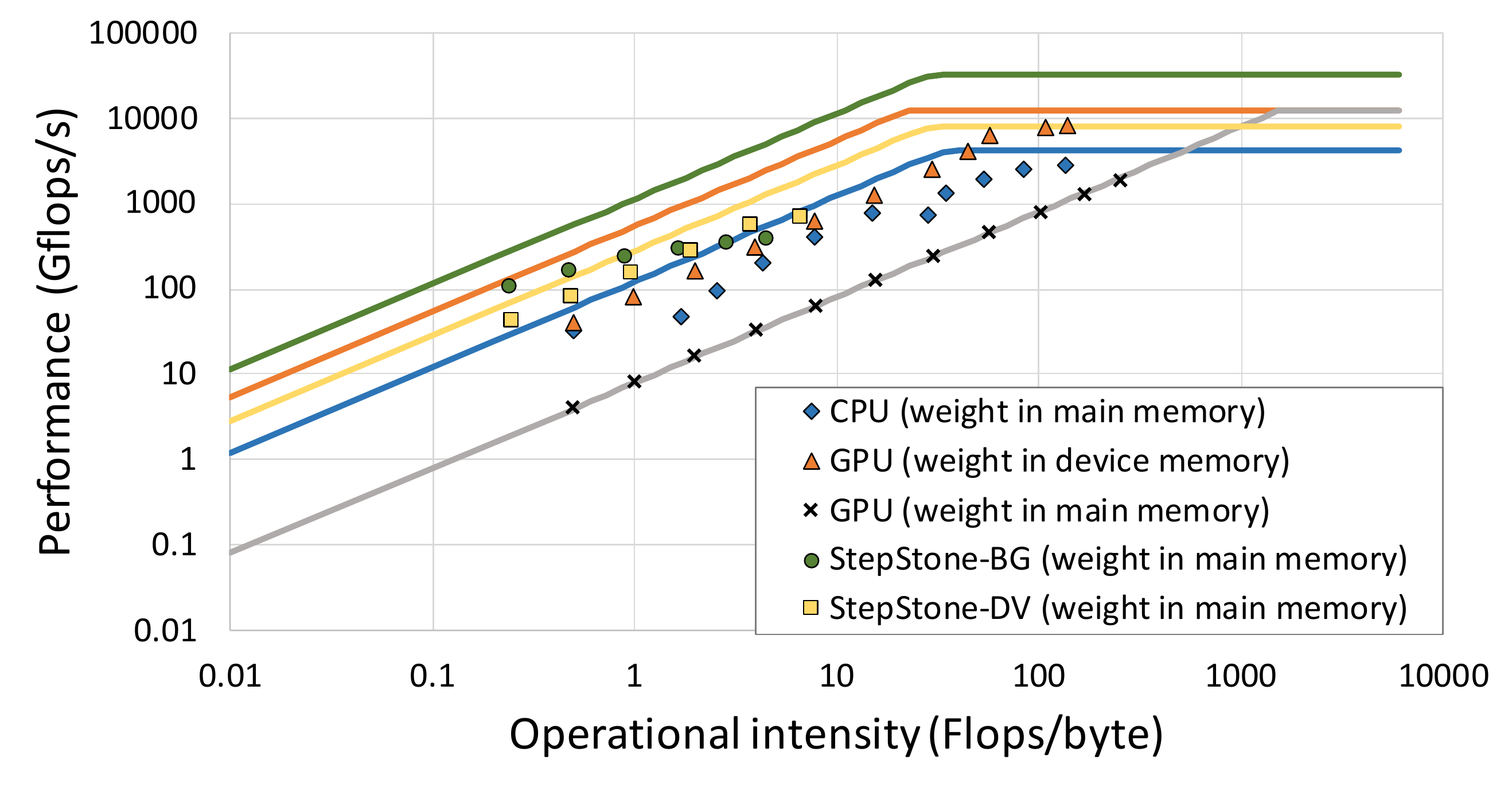}
	\caption{Roofline models for CPU, GPU, and StepStone PIMs; measured results are for a 1K $\times$ 4K weight matrix for varying batch sizes (the left most point of each system is for batch-1 and the batch is 2$\times$ larger for each point moving to the right).}
	\label{fig:eval_roofline_1k_4k}
	\vspace*{-3mm}
\end{figure}

\noindent\textbf{Throughput rooflines.} 
\label{subsec:eval_throughput}
The throughput benefits of StepStone are also apparent on the roofline plot presented in \fig{fig:eval_roofline_1k_4k}, also for a 1024 $\times$ 4096 weight matrix. The plot includes the CPU as above, StepStone-BG and -DV (the maximum of the two represents the achieved performance with StepStone), and the performance obtained with an NVIDIA Titan Xp (running CUTLASS) when the model is already present in GPU device memory or when it must first be read from CPU main memory.

In the realistic scenario where GPU memory capacity is too small to accommodate the full recommender system and language models, StepStone exhibits higher throughput (in addition to its latency benefits) at all reasonable batch sizes. In fact, the CPU and GPU offer an advantage only once the batch is 256 samples or greater. Even if the model fits in GPU memory, StepStone offers higher performance for batches of 16 samples or less.
The gap between the rooflines and simulated performance of StepStone stems from the localization and reduction overheads.


 We emphasize that StepStone PIM achieves this high performance benefits  without utilizing CPU or GPU compute resource, such as cores or caches. This implies that the overall system performance can increase even further by colocating additional tasks on the same node.

\subsection{End-to-End Performance}
\label{subsec:eval_end_to_end}

\begin{figure}[t!]
\centering
        \includegraphics[width=0.48\textwidth]{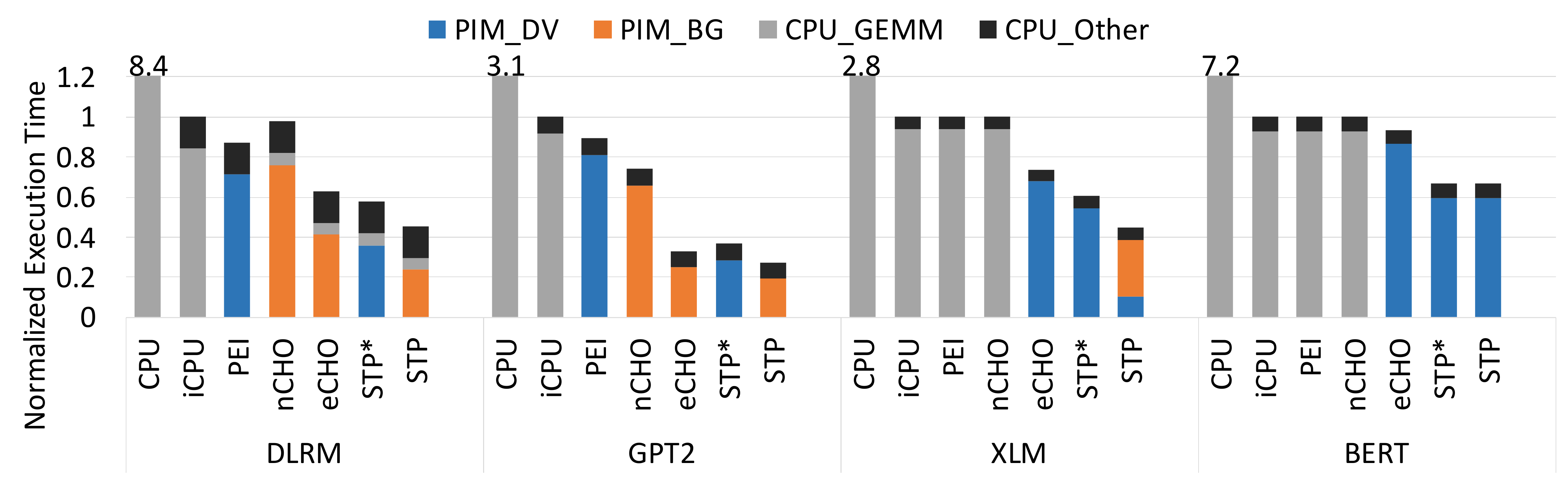}
	\caption{End-to-end performance results for various recommendation and language models with the CPU and PIMs.}
	\label{fig:end_to_end}
	\vspace*{-3mm}
\end{figure}

\fig{fig:end_to_end} compares the inference performance of one recommendation system and three language models with different PIM approaches---PEI, Chopim, and StepStone PIM---to that of a CPU. For the PIM approaches, we assume the same PIM system (\fig{fig:baseline_system}) and that GEMMs can be executed by either the CPU, device-level (PIM\_DV), or BG-level PIMs (PIM\_BG); the best performing option is chosen for each GEMM. GEMMs are used for FC and projection layers. All other operations, including concatenation, GELU, softmax, sorting, batched GEMM, and some data reorganization (e.g. stack) operations are executed on the CPU (\textit{CPU\_Other}). 

We show the performance of two different CPU models: measured on the real system (\textit{CPU}) and idealized CPU performance (\textit{iCPU}). We estimate idealized performance with our StepStone-CH, which maximally utilizes memory channel bandwidth. Overall, the measured results show that the CPU performs poorly on small-matrix GEMMs. 

Naive Chopim (\textit{nCHO}) executes GEMMs as multiple GEMV operations, which leads to missed locality opportunities. On the other hand, if Chopim is enhanced with StepStone block grouping (\textit{eCHO}) and divides each GEMM into smaller dot-product operations, it benefits from better PIM buffer locality and the overhead for buffer fill/drain significantly decreases. However, compared to StepStone PIM, eCHO suffers from higher localization/reduction overhead. We evaluate a low-power StepStone PIM mode (\textit{STP*}), where only StepStone-DV is used, and a high-performance mode (\textit{STP}), which selects the best-performing level per GEMM. 


The execution time of DLRM is dominated by a single FC layer (92\%) and GEMM execution time is long enough to amortize the localization/reduction overheads. This enables Chopim and StepStone PIM to use BG-level PIMs and benefit from their high memory bandwidth. On the other hand, PEI cannot fully utilize BG-level PIMs due to command bandwidth bottleneck and, consequently, using more PIMs with PEI only increases overhead. GPT2 shows a similar trend but the gaps between PEI and StepStone PIM/Chopim are greater due to a larger weight matrix than DLRM. 
In BERT and XLM, the $N$ dimension is the batch size multiplied by the sequence length after tensor reshaping, offering more efficient GEMM execution. For BERT, \textit{N} becomes 32 in all FC layers whereas, for XLM, the sequence length starts at 1 and increases by 1 up to the maximum length (8 in our configuration) after each iteration. As a result, XLM utilizes BG-level PIMs when \textit{N} is small and, later, switches to DV-level PIMs once arithmetic performance saturates and overheads start to dominate.

Overall, when weight matrices are larger and the batch dimension is smaller, StepStone PIM outperforms other CPU and PIM approaches. \meadd{Even with somewhat larger batches (e.g., up to $N=384$ for BERT), StepStone PIM outperforms the CPU by splitting a batch into several batch-32 GEMM operations. For example, StepStone PIM achieves 12${\times}$ higher performance than the CPU for BERT. Thus, StepStone PIM outperforms the CPU until $N=12\times32=384$.}  

\subsection{Impact of StepStone AGEN}
\label{subsec:eval_agen}

\begin{figure}[t!]
\centering
        \includegraphics[width=0.50\textwidth]{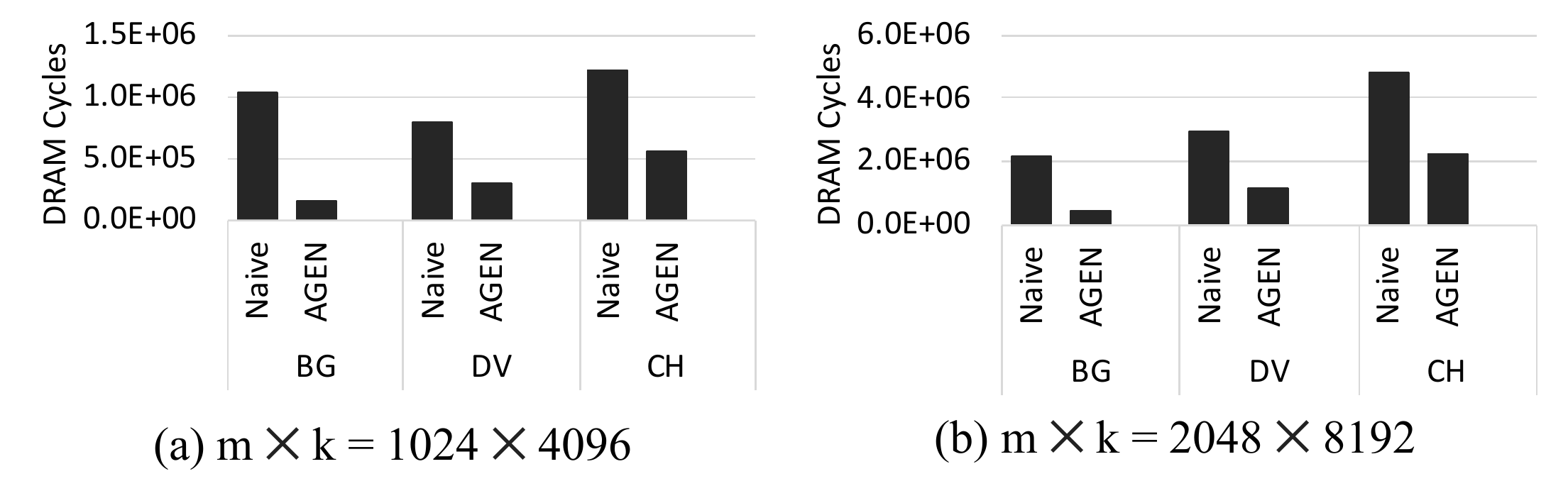}
	\caption{GEMM latency comparison between naive address generator and the proposed StepStone AGEN.}
	\label{fig:eval_agen}
	\vspace*{-2mm}
\end{figure}

\fig{fig:eval_agen} shows the performance benefit of StepStone AGEN over the naive approach (explained in Section~\ref{subsec:gemm_addr_gen}). Overall, StepStone AGEN outperforms the naive approach by up to 4$\times$, in particular when the number of active PIMs is larger. Intuitively, the naive approach can find the next cache block in a probability of ${1/n}$, where ${n}$ is the number active PIMs. For StepStone-BG (\fig{fig:eval_agen}{a}) there are 16 active PIMs and the performance difference between two approaches is 8${\times}$. This is mainly because pairs of cache blocks are contiguous in our baseline address mapping, which equates the naive approach with our optimized AGEN. However, when a large gap in the mapping exists, the naive approach requires numerous iterations and requires a large number of cycles to generate the next address. The DRAM burst transfer latency is 4 DRAM cycles and bubbles are introduced any time the next address cannot be generated within that time window. This does not occur with our proposed AGEN and its latency can always be hidden within the pipeline. The difference in performance between the two approaches for this case is apparent for StepStone-DV with a large weight matrix (\fig{fig:eval_agen}{b}), where the performance gap is 2.5${\times}$.

\subsection{Parallelism---Distribution Overhead Tradeoffs}
\label{subsec:eval_tradeoff}

\begin{figure}[t!]
\centering
        \includegraphics[width=0.48\textwidth]{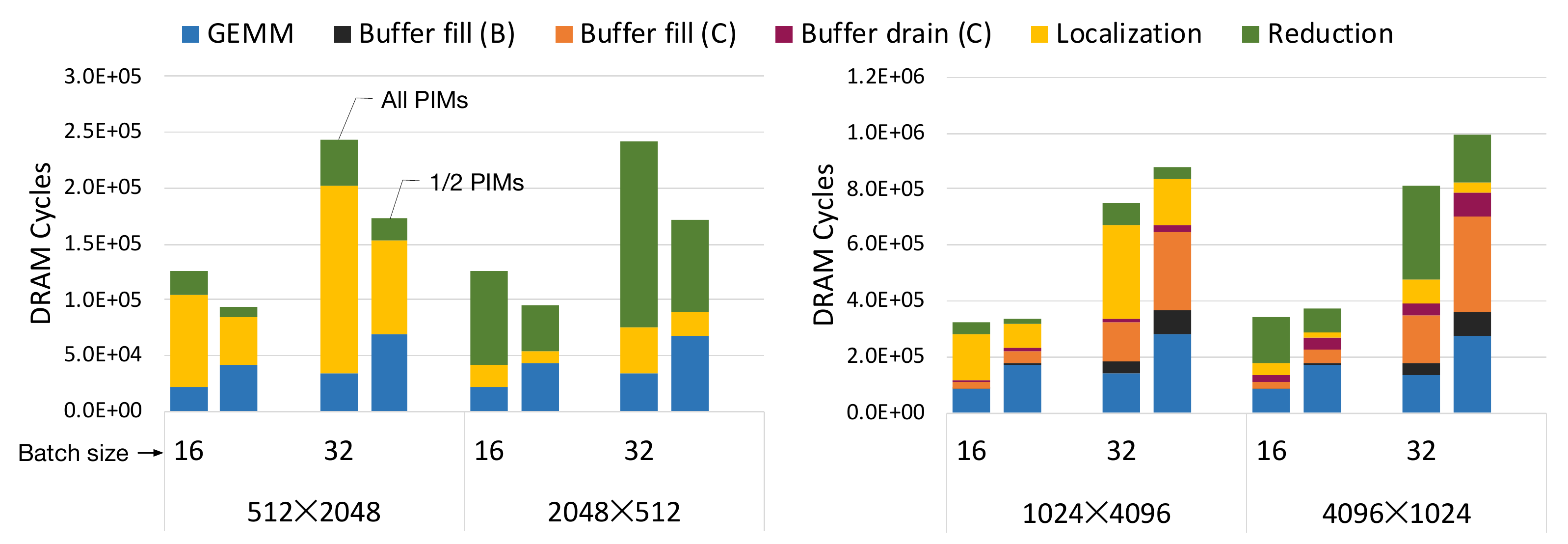}
	\caption{Impact of trading off between PIM execution time and replication/reduction overhead.}
	\label{fig:eval_tradeoff}
	\vspace*{-3mm}
\end{figure}

\fig{fig:eval_tradeoff} shows the GEMM latency comparison between two cases: (1) when all bank group-level PIMs are used and (2) only half of the BG-level PIMs are active. We present bank group-level PIMs tradeoff because we already discussed tradeoffs with respect to PIM level. When the weight matrix size is small, the fraction of replication and reduction overhead dominates the entire execution time. If we only activate half of the BG-level PIMs the overheads decrease by at most half while arithmetic execution time doubles because parallelism is cut in half. Still, the tradeoff proves beneficial when the matrices are small (left). On the other hand, as the matrix size increases, the fraction of PIM execution time increases as well (right). This is because the PIM execution time quadruples as each dimension size is doubled, whereas the localization/reduction overhead only doubles. Moreover, as the input and output matrices (i.e. activations) grow, they exceed scratchpad capacity. As a result, the fraction of execution time required for buffer fill/drain operations also increases. Even though using fewer PIM units does offer better performance for the larger matrix, it still provides a valid tradeoff option because comparable performance is attainable in some cases while resource usage and power consumption decreases.

\subsection{Impact of Address Mapping}
\label{subsec:eval_addr_map}

\begin{figure}[t!]
\centering
        \includegraphics[width=0.48\textwidth]{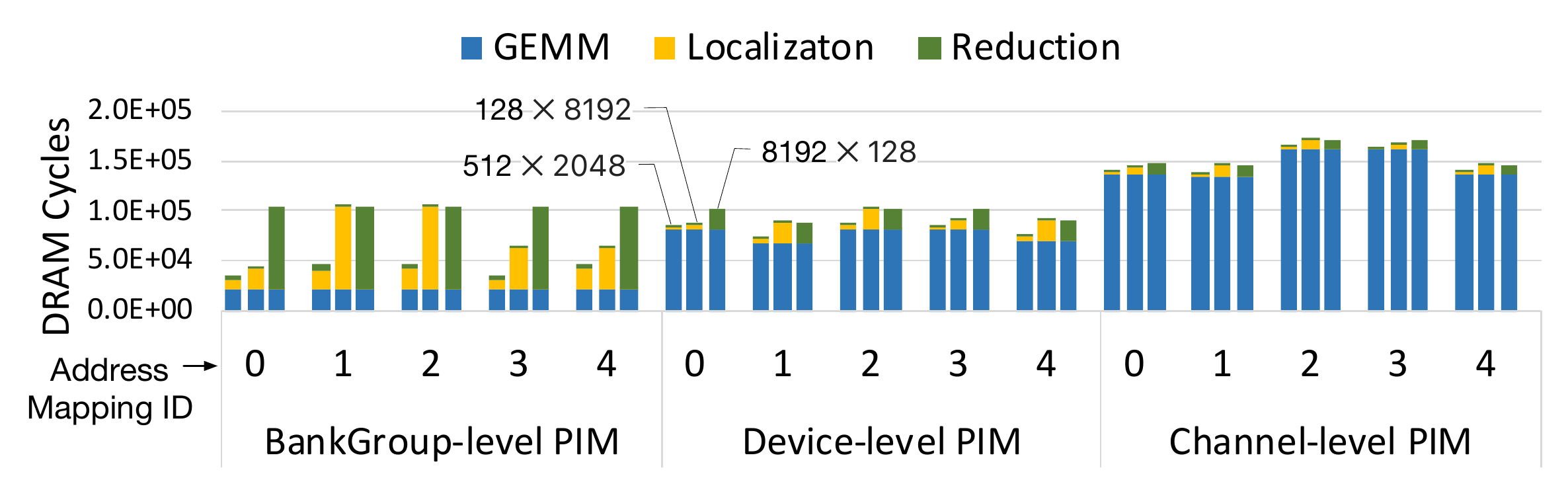}
	\caption{Sensitivity to address mapping and aspect ratio of the weight matrix (batch\_size = 4).}
	\label{fig:eval_addr_map}
	\vspace*{-3mm}
\end{figure}

\fig{fig:eval_addr_map} shows the execution time of GEMM operations when different address mappings and aspect ratios of the weight matrices are used. To isolate impact to only DRAM address mapping, we set the batch size to 4 such that the input and output matrices fit in the scratchpad memory of all three PIM options. In the bank-group level StepStone-BG, the fraction of localization overhead differs significantly across address mappings when the matrix is short and fat (i.e., 128 ${\times}$ 8192). This is because the number of PIMs that share the same input matrix blocks in address mappings 1 and 2 are 2${\times}$ greater than those with address mappings 3 and 4 and 4${\times}$ greater than those with address mapping 0. The reason for the low localization overhead with address mapping 0 is that the combination of the address mapping and matrix size interleaves addresses such that matrix columns remain contiguous within each PIM.
In contrast, the tall and thin GEMM (i.e., 8192 ${\times}$ 128) suffers from high reduction overhead for all address mappings. This is because  the CPU address mappings choose fairly fine-grain interleaving across bank groups and channels to maximize bandwidth. StepStone-BG is more sensitive to address mapping and aspect ratio compared to StepStone-DV and -CH, because it distributes work across a larger number of units and the relative overhead of localization and reduction is higher.

Note that address mappings 2 and 3 for StepStone-CH exhibit higher GEMM execution times because these mappings interleave bank groups at a coarser granularity. Hence, bandwidth cannot be maximized because consecutive  memory accesses are penalized by tCCDL (i.e., column-to-column delay for back-to-back accesses within the same bank group is larger than across bank groups). This demonstrates that timing parameter considerations are also important when deciding the address mappings for PIM-enabled main memory. In theory, the  localization and reduction overheads are lower when fewer PIMs share the input and output matrix blocks as common operands. However, this goal of low sharing cannot be realized with a single fixed address mapping because the sharing pattern changes with the matrix size and aspect ratio. 

\subsection{Impact of Scratchpad Memory Capacity}
\label{subsec:buffer_size}

\begin{figure}[t!]
\centering
        \includegraphics[width=0.48\textwidth]{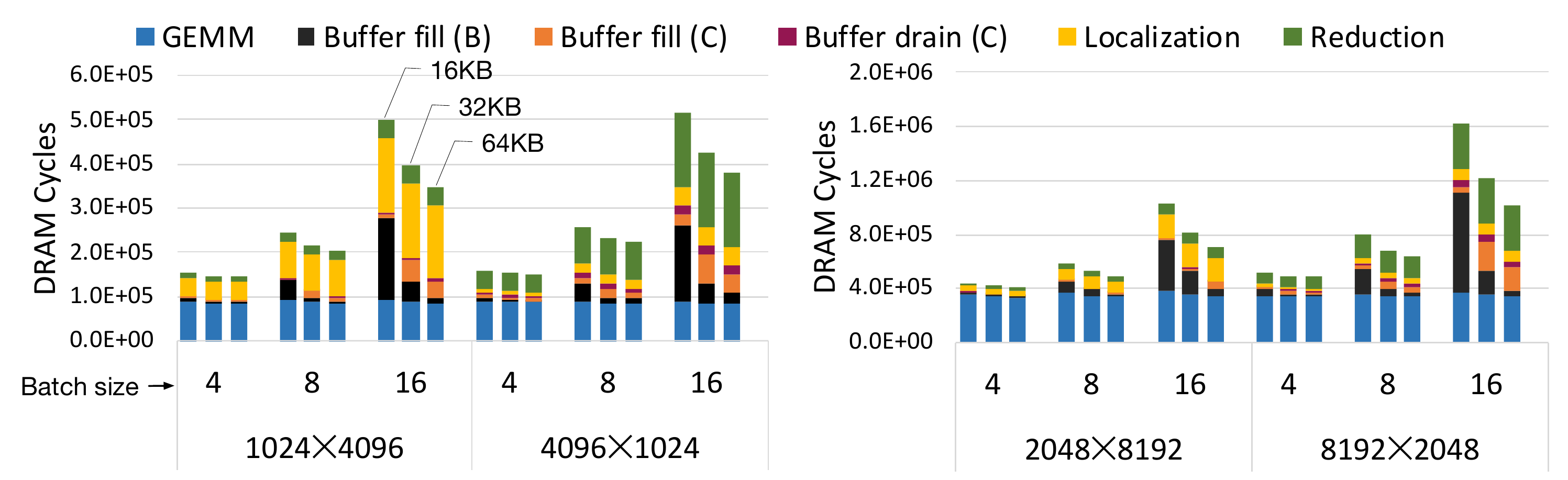}
	\caption{GEMM latencies for different matrices and buffer sizes (StepStone-BG).}
	\label{fig:eval_buffer_size}
	\vspace*{-3mm}
\end{figure}

\fig{fig:eval_buffer_size} shows the impact of scratchpad memory capacity on GEMM latency. We analyze StepStone-BG as it has the most stringent area constraint. We search for an optimal allocation across the scratchpad  partitioning options between input and output buffer allocations (there are only two buffers so the search converges quickly). We find that  interleaving buffer fill/drain operations with arithmetic has negligible impact on GEMM performance. The ability to execute entire kernels limits the benefits of overlapping data transfer with arithmetic and interleaving increases the number of row buffer conflicts, though row-buffer locality remains high.

Larger matrices (right) tends to amortize the buffer fill/drain overheads better than smaller matrices (left). Generally, overhead increases with batch size. Interestingly, the overhead with the 2048 $\times$ 8192 weight matrix increases at half the rate of other matrix configurations. This is because the number of block groups with this specific weight matrix dimension is half that of the other matrix sizes we evaluate. Consequently, the working set  of the input activation (matrix B) per PIM unit is also half that of other matrix configurations. As explained in Section~\ref{subsec:gemm_grouping}, the number of groups is determined by both the address mapping and matrix dimensions. 

\subsection{Impact of Concurrent CPU Access}

\meadd{
We expect StepStone PIM to outperform prior PIM architectures, including Chopim, by enabling longer-running GEMM operations that maintain PIM locality. Long-running operations are important when the CPU also executes a memory-intensive workload concurrently with the PIMs, as both the CPU and PIMs contend for limited command channel bandwidth. We evaluate this using the same colocation used by Cho et al.\ for evaluating Chopim~\cite{cho2020near}, as described in Section \ref{sec:method}. While the colocated applications are not DL-related, they run readily on gem5 and clearly demonstrate the impact of command channel contention. We isolate the performance benefits to just the StepStone AGEN unit that enables long-running kernels by running the same StepStone GEMM flow on eCHO and StepStone PIM and reporting results corresponding only to GEMM execution.}

\begin{figure}[t!]
\centering
        \includegraphics[width=0.48\textwidth]{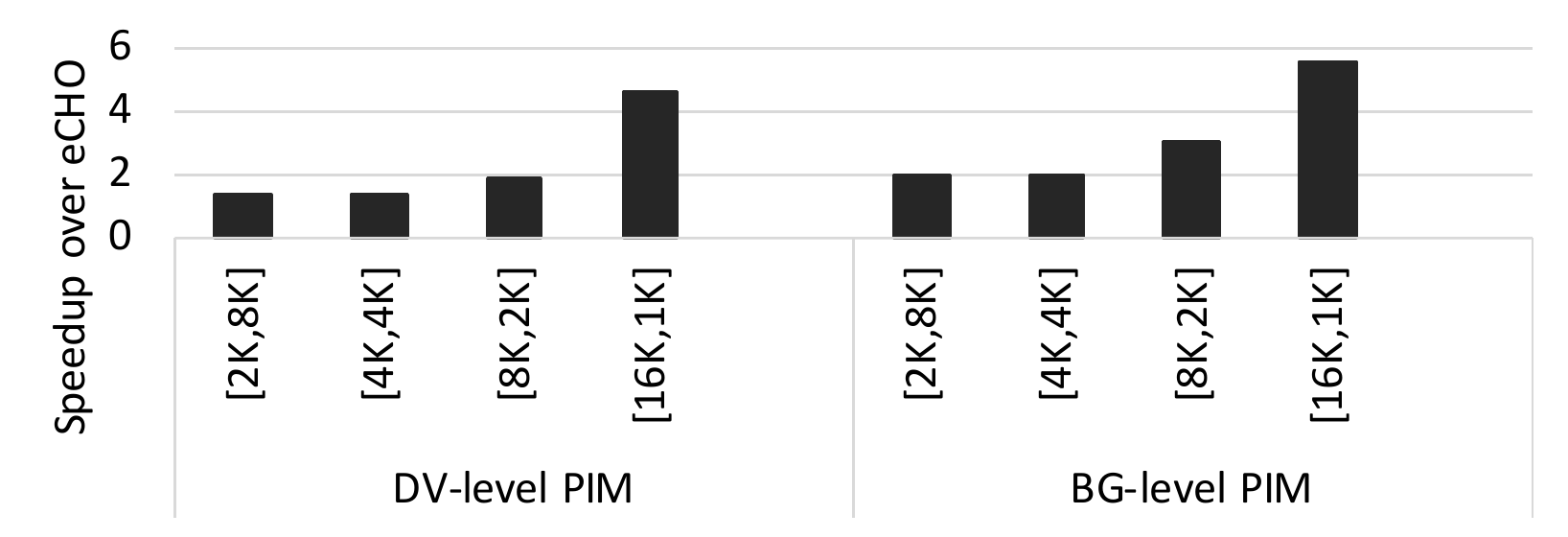}
	\caption{Speedup of StepStone PIM (STP) over Chopim enhanced with StepStone block grouping (eCHO) when concurrent CPU access exists. The size of matrices is fixed and its aspect ratio is varied.}
	\label{fig:concurrent_access}
	\vspace*{-2mm}
\end{figure}

StepStone PIM outperforms Chopim when the CPU intensively accesses memory concurrently with PIMs~(\fig{fig:concurrent_access}). As the matrix shape changes from short-fat to tall-thin, each of eCHO kernels accesses fewer cache blocks, resulting in more PIM kernel invocations and greater contention with the CPU for the command channel. As a result, PIM kernel packets are delayed and the PIMs are underutilized. With BG-level PIM, the relative performance of Chopim to StepStone PIM is worse since even more PIMs are underutilized due to the command bandwidth bottleneck. This performance gap will increase as the number of PIMs in each channel increases, increasing the importance of mechanisms that enable long-running kernels. 

\subsection{Power and Energy Analysis}
\label{subsec:eval_power}

\begin{figure}[t!]
\centering
        \includegraphics[width=0.48\textwidth]{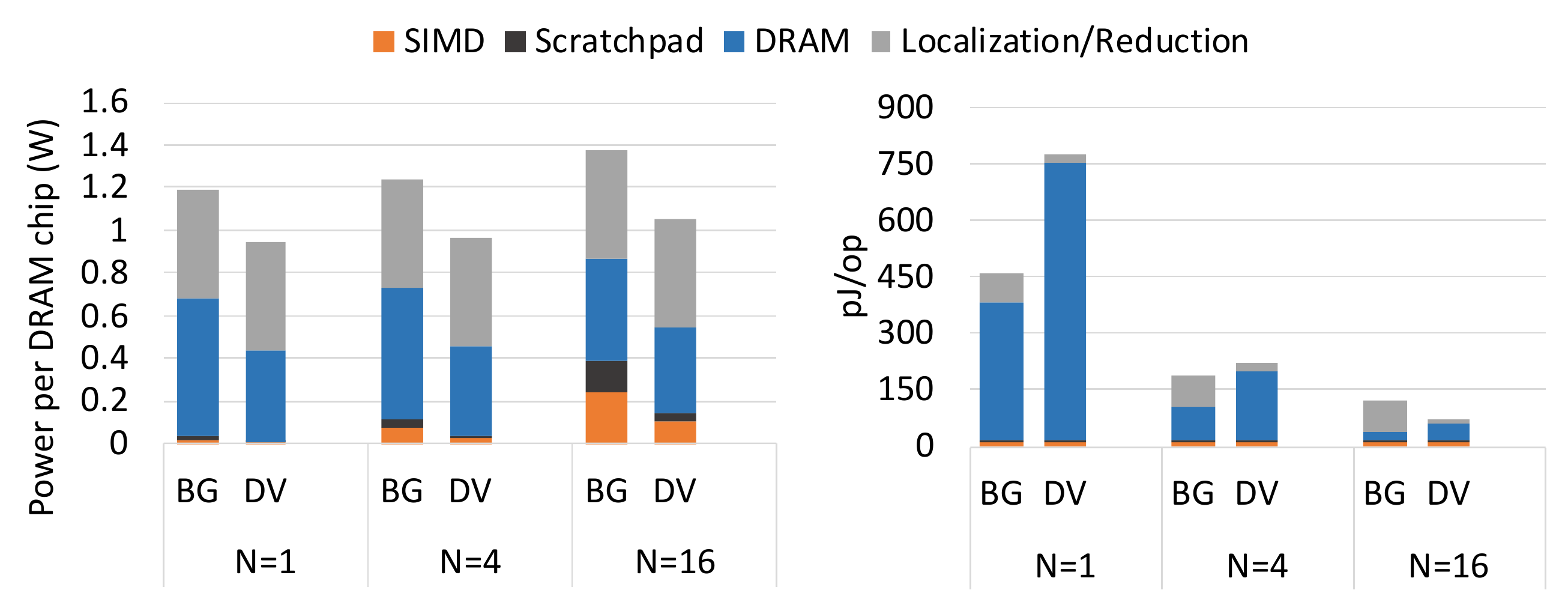}
	\caption{Power dissipation per DRAM device (left) and energy consumption per floating-point operation (right) of StepStone-BG and StepStone-DV (weight\_matrix = [1024, 4096]).}
	\label{fig:eval_power}
	\vspace*{-2mm}
\end{figure}

\fig{fig:eval_power} shows the power and energy consumption per DRAM device of StepStone-BG and StepStone-DV. As ${N}$ increases, the relative contribution of arithmetic also increases. However, overall, the power of DRAM access (either within the PIMs or for localization and reduction) dominates the power of the SIMD units. The right side of the figure shows that StepStone-BG is more energy-efficient than StepStone-DV when ${N}$ is small. The main source of this energy savings is that IO energy is much smaller within a device. However, as ${N}$ increases, the energy for localization and reduction dominates the energy consumption of arithmetic and StepStone-DV is more efficient. Note that, if power exceeds the delivery/cooling budget for a chip or module, performance can be throttled.

\section{Related Work}
\label{sec:related_work}

To the best of our knowledge, this is the first work that enables high-performance and CPU-compatible near-memory processing for accelerating bandwidth-bound fully-connected (FC) layers. We describe prior related work below and contrast our contributions for existing approaches. 


\medskip
\noindent\textbf{\textit{Processing in main memory.}}
Processing in main memory implies that PIM should play along with the other parts of the system; otherwise, it will have a system-wide impact. Considering this, some researches \cite{ahn2015pim,kim2017toward} enable PIM in a fine granularity, such as PIM operations per cache block. This approach can solve the complex address mapping problem. The CPU indicates the next cache block to process with some command packets and PIM processes the cache block. Though this approach can accommodate more applications due to its flexibility, PIM performance will be eventually limited by the command bandwidth. RecNMP \cite{ke2020recnmp} mitigates this command bandwidth bottleneck by sending compound of memory requests but this solution does not scale when there are more than 4 PIMs per channel. Chopim \cite{cho2020near} enables coarse-grained PIM kernels under complex address mapping. Though GEMM can be executed with Chopim by multiple GEMV kernel calls, temporal locality cannot be exploited which is crucial for high-performance GEMM operations. Also, Chopim does not provide efficient localization and reduction mechanisms, which incur high overhead for executing GEMMs on PIMs. NDA \cite{farmahini2015nda}, Chameleon \cite{asghari2016chameleon}, and MCN DIMM \cite{alian2018nmp} are also based on conventional DIMM devices and proposes PIM designs to practically add PEs to main memory. 


\medskip
\noindent\textbf{\textit{GEMM acceleration with PIM.}}
The Active Memory Cube (AMC) \cite{sura2015data} targets GEMM operations with in-memory processors. AMC considers address interleaving but interleaving is only allowed within the same local PIM memory. This essentially requires partitioning into CPU and PIM memory spaces and data should be copied from one space to another if the processing mode changes. This approach has the same data loading problem as discrete accelerators and does not truly realize PIM potential of sharing the same memory between the CPU and PIM. On the other hand, our solution does not have this limitations and works with any XOR-based address mapping and PIMs in any DRAM hierarchy levels. 

\medskip
\noindent\textbf{\textit{PIM for machine learning.}}
PIM for machine learning workloads has been widely studied. Much of this research targets convolutional neural networks~\cite{long2019design,imani2019floatpim,gao2017tetris,chi2016prime,deng2019lacc, joardar2019regent, kim2016neurocube, shafiee2016isaac, song2017pipelayer}, embedding table lookups in recommendation models~\cite{ke2020recnmp, kwon2019tensordimm}, recurrent neural networks~\cite{long2018reram}, and GAN~\cite{mao2018lergan,rakin2018pim}. In contrast, we target the tall-thin/fat-narrow GEMMs of fully-connected layers in DL inference. Newton~\cite{he2020newton} also targets fully-connected layers, like StepStone PIM. However, Newton operates as a discrete accelerator that cannot benefit from the advantages of main-memory acceleration described in Section~\ref{sec:motiv}. More importantly, Newton does not avoid weight copies, does not exploit GEMM locality, cannot trade off parallelization degree overheads with performance benefits, cannot selectively execute at different PIM levels or the CPU to dynamically match changing workload characteristics, and does not support the long-running kernels necessary for concurrent bandwidth-intensive CPU tasks.


\section{Conclusion}
\label{sec:coclusion}
We identify that small-batch GEMM operations of DL inference workloads are bandwidth bound on CPUs and GPUs and can benefit from PIM acceleration. We introduce \textit{StepStone PIM}, which enables independent PIM GEMM execution under complex CPU DRAM address mapping. The main innovation of StepStone PIM is the address-mapping cognizant GEMM blocking with matching PIM-side address generation. Our unique AGEN logic improves throughput compared to naive or host-side address generation. We explore PIM designs in three different DRAM hierarchy levels (channel, chip, and bank-group levels) and show their tradeoffs with detailed simulation results. We show that activating more PIMs for GEMM improves arithmetic performance but adds overheads for data localization/replication and reduction. 
We conclude that StepStone is an appealing datacenter solution because of: (1) its low cost; (2) its potential for lower latency and higher throughput even when implemented at the buffer-chip level within a DIMM without DRAM device modification; and (3) its locality-optimized high efficiency GEMM execution that frees CPU resources for other tasks. 



\bibliographystyle{IEEEtranS}
\bibliography{reference.bib}

\end{document}